\documentclass[10pt]{article}

\usepackage[pdftex]{graphicx}
\graphicspath{{./pdf/}}

\usepackage{csagh}
\usepackage[utf8]{inputenc}   
\usepackage[T1]{fontenc}         
\usepackage[english]{babel} 
\usepackage{booktabs}            
\usepackage{hyperref}

\begin{document}
\begin{opening}

\title{Parallel Memory Management in Goose, C++, and Rust: An Experimental Study of Allocation, Reclamation, and Scaling}

\author[AGH University of Krakow, al. Mickiewicza 30, 30-059 Cracow, rafalfr@agh.edu.pl]{Rafa\l{} Fr\k{a}czek}

\begin{abstract}
Compiler-managed memory can reduce allocation and reclamation overhead, but its behavior under parallel workloads remains insufficiently characterized. This study evaluates Goose v2026.3 against C++ and Rust implementations, including tree variants with individually allocated nodes and contiguous arenas. We benchmarked worker-local dynamic arrays and binary trees using multiple workers on a workstation with four physical cores and eight logical CPUs. Repeated tree experiments included five construction, traversal, and reclamation rounds. Each main experimental configuration was measured in 15 independent process runs. We measured whole-process execution time, release-operation intervals, and memory usage at synchronized checkpoints. We also examined strong and weak scaling. For depth-20 trees with eight workers, Goose achieved 5.62\% and 10.95\% lower median execution times than the C++ and Rust arenas, respectively. Its additional process RSS at the LIVE checkpoint was comparable to the arenas and substantially lower than that of individually allocated trees, but its virtual memory size was much larger. Logical reclamation did not always reduce resident memory, and Goose showed lower weak-scaling efficiency than the C++ and Rust arenas. Results for dynamic arrays were mixed. These results indicate that compiler-managed storage can be effective for parallel workloads with predictable object lifetimes. However, execution time alone does not capture its memory and scalability trade-offs.
\end{abstract}


\end{opening}

\section{Motivation}

In modern computer architectures, memory allocation and reclamation can affect the performance and scalability of parallel programs, especially when many objects are created and released. Previous studies have shown that allocation policies influence execution time and memory consumption \cite{DBLP:conf/asplos/BergerMBW00} \cite{DBLP:conf/oopsla/BergerZM02}. The new programming language Goose uses compiler-managed data stacks and scope-based reclamation \cite{Wouters2026}. However, existing Goose benchmarks have focused mainly on single-threaded workloads \cite{bakker2026goosebench}. Its behavior under parallel allocation and reclamation remains insufficiently characterized.
C++ uses RAII, smart pointers, and standard containers to manage resources and object lifetimes \cite{DBLP:books/daglib/0019344}. Rust uses ownership and borrowing to provide statically checked memory-safety guarantees for safe code without a garbage collector \cite{DBLP:journals/pacmpl/0002JKD18}. Both languages support individual object allocations and arena-based storage. These approaches differ in allocation granularity, data representation, and reclamation behavior.

In Goose programming language, memory allocation advances the data-stack pointer. Restoring a saved pointer makes the allocated storage reusable. Goose also supports compact relative references and worker-local data stacks \cite{Wouters2026}. Region-based allocation and bulk reclamation are not new techniques \cite{DBLP:journals/iandc/TofteT97}. What matters here is how Goose implements these techniques and how they perform under parallel workloads. Previous Goose benchmarks \cite{bakker2026goosebench} do not fully address parallel scaling, repeated reclamation, or post-release memory usage. Furthermore, logical reclamation does not necessarily return physical memory to the operating system.

This study compares Goose v2026.3 with C++ and Rust using three parallel workloads: dynamic arrays, binary trees, and repeated tree allocation and reclamation. The C++ and Rust tree implementations include individually allocated nodes and optimized contiguous arenas. We measure whole-process execution time, memory usage, and release-operation duration. Strong and weak scaling are also evaluated. The objective is to identify the benefits and limitations of managed storage in different workloads. We do not aim to establish that one programming language is universally superior.

The structure of the paper consists of the following sections. In section 2, relevant techniques will be reviewed and research gaps will be outlined. Research questions and contributions are described in section 3. Experimental methodology including execution environment, benchmarks and procedures will be presented in section 4. Section 5 contains a comparative discussion of the experimental results. The main findings and limitations are highlighted in section 6. The conclusion will follow in the final section 7.

\section{Introduction}

\subsection{State of the Art in Memory Management}

Memory management methods affect execution time, process memory usage, and parallel scalability. Memory allocation methods balance flexibility, allocation costs, and memory locality. Specialized policies do not always outperform general-purpose allocators \cite{DBLP:conf/oopsla/BergerZM02}. C++ uses RAII and smart pointers to manage resource lifetimes \cite{DBLP:books/daglib/0019344}. Rust uses ownership and borrowing to provide memory-safety guarantees. The RustBelt project established safety results for modeled Rust mechanisms \cite{DBLP:journals/pacmpl/0002JKD18}. However, lifetime management does not determine the allocation strategy or data representation. Both languages support individual allocations and containers with contiguous virtual memory addresses. This does not imply contiguous physical memory.

Region-based memory management algorithms group objects with compatible lifetimes and release them together. Tofte and Talpin \cite{DBLP:journals/iandc/TofteT97} developed its formal foundations. Grossman et al. \cite{DBLP:conf/pldi/GrossmanMJHWC02} introduced region-based allocation in Cyclone, with static safety guarantees and lifetime controls. These approaches can reduce individual deallocation costs. However, they may retain objects longer and impose restrictions on object lifetimes.

Memory pools and arenas provide related allocation strategies. Lattner and Adve \cite{DBLP:conf/pldi/LattnerA05} studied the allocation and data layout driven by compilers. Compact layouts can reduce allocation overhead and improve memory locality. However, collective reclamation limits the ability to release individual objects.

Parallel memory management introduces additional challenges, including synchronization, allocator contention, and cross-thread deallocation. The Hoard allocator addresses memory utilization and scalability in multithreaded programs \cite{DBLP:conf/asplos/BergerMBW00}. Evans introduced jemalloc for scalable concurrent allocation \cite{Evans2006jemalloc}. Jemalloc uses multiple arenas to reduce contention. The mimalloc allocator uses page-local free-list sharding to improve allocation locality \cite{DBLP:conf/aplas/LeijenZM19}. Hazard pointers support safe reclamation of shared objects \cite{DBLP:journals/tpds/Michael04}. In contrast, our benchmarks use worker-local arrays and binary trees. Their elements and nodes are not shared between workers, although synchronization and communication are still required.

Recent studies also highlight the importance of workload characteristics. Zhou et al. \cite{DBLP:conf/asplos/ZhouGVK0KMDR24} examined the effects of object sizes and lifetimes on TCMalloc performance. Hughes and Tratt \cite{DBLP:journals/pacmpl/HughesT25} investigated garbage collection in Rust, including a binary-tree benchmark based on an arena. He et al. \cite{DBLP:journals/pacmpl/HeJBR26} proposed a type system that supports more flexible arena lifetimes. These studies show that allocation performance depends on data representation, object lifetimes, and workload characteristics.

Existing research addresses different aspects of memory management, but no single strategy is optimal for every workload. The studies reviewed here do not directly compare Goose's worker-local storage with optimized C++ and Rust arenas under parallel execution. In particular, the relationship between logical reclamation, process memory usage, and parallel scaling remains insufficiently characterized for this comparison.

\subsection{Research Gap and Rationale}

The existing Goose benchmark suite compares several data structures implemented in Goose, C++, and Rust \cite{bakker2026goosebench}. Its binary-tree benchmarks include individually allocated nodes and optimized arenas. They also examine repeated tree construction and destruction. However, these benchmarks focus on single-threaded execution time and peak memory consumption. They do not establish how performance and memory usage change with worker count. They also do not measure memory at synchronized points during repeated allocation and reclamation.

Three questions remain. First, how does Goose compare with optimized C++ and Rust storage under parallel workloads? Second, how does logical release affect process memory usage? In particular, how do resident memory and virtual address space differ? Third, how does performance change under strong and weak scaling?

We study these questions with dynamic arrays, binary trees, and repeated tree construction and release. We measure process execution time, release duration, and memory at synchronized checkpoints. We aim to identify trade-offs, not to establish that one language is always faster.

\section{Goals of the research}
\subsection{Research Questions}

The purpose of this study is to evaluate the performance and memory management behavior of Goose v2026.3. The comparison is conducted between Goose, C++, and Rust implementations for workloads that include parallel computations. This research focuses on memory allocation methods, object lifetime issues, and the impact of concurrent computations. In order to define the directions for conducting research work, the research questions listed below are formulated.

Question 1. What are the differences between Goose's memory storage and approaches using object-by-object and arena allocation techniques in terms of execution time and memory usage?

The research question addresses computational and memory costs related to various ways of handling dynamic data structures. Especially, it distinguishes between individual objects allocation and arena based memory management. The experiments include dynamic arrays and binary trees. Both dynamic arrays and binary trees are built with concurrent computation capabilities.

Question 2. How do release durations differ between implementations? How does logical reclamation affect the process resident set size (RSS)?

This question examines the cost of logical memory reclamation. We also investigate whether reclamation reduces process RSS. The measurements cover memory usage before allocation, while objects remain allocated, and after logical reclamation. We compare these results across implementations.

Question 3. How does execution time change with worker count under strong and weak scaling?

This question examines how execution time changes with the number of workers. Strong scaling keeps the total workload constant, while weak scaling increases the total workload in proportion to the number of workers. Both scenarios are evaluated using parallel binary-tree workloads.

\subsection{Contributions}

\textbf{Benchmark comparison}. We extend previous Goose benchmarks \cite{bakker2026goosebench} to parallel workloads and repeated memory reclamation. We compare Goose with two C++ and Rust tree representations. The first uses separately allocated nodes managed by \texttt{unique\_ptr} in C++ or \texttt{Box} in Rust. The second stores compact nodes in a \texttt{vector} or \texttt{Vec} arena.

\textbf{Memory measurements}. We use four synchronized checkpoints: BASE before tree construction, LIVE while trees are allocated, ROUND after release, and FINAL after worker termination. We measure process RSS and virtual memory size (VMS). Logical release does not necessarily reduce resident memory.

\textbf{Release and scaling measurements}. We measure construction and traversal together. We time release operations separately and exclude synchronization waiting from these intervals. We also evaluate strong and weak scaling.

\textbf{Reproducibility}. The main experiments use 15 independent process executions per configuration. We verify checksums and node counts in all implementations. We preserve source files, execution scripts, and measurement records to support the reproduction of the results.

\section{Experimental Methodology}

\subsection{Experimental Environment and Implementation}

Experiments have been performed using the hardware configuration consisting of the Intel Core i3-14100 CPU with four physical cores and eight logical processors and 32 GiB of RAM. It should be noted that the OS used is Linux with the 7.2.9 kernel version.

The experiments used Goose v2026.3, GCC 16.2.1, and Rust 1.98.0 (edition 2021). Goose and C++ were compiled with \texttt{-O2}, while Rust used \texttt{-C opt-level=2}. The repeated reclamation benchmark used C++20. The dynamic array and standalone tree benchmarks used C++17. The C++ builds also used \texttt{-DNDEBUG} and \texttt{-pthread}. Both tree benchmarks used the Rust option \texttt{-C codegen-units=1}. The dynamic array benchmark did not specify this option. The repeated reclamation metadata recorded the Goose release and the SHA-256 hash of the Goose compiler executable.

The experiments used 1, 2, 4, and 8 workers. For configurations with up to four workers, the affinity of the process was restricted to logical CPUs 0, 2, 4 and 6. These logical CPUs belong to four different physical cores. The eight-worker configuration could use all eight logical CPUs. Individual workers were not tied to specific cores. Thread placement was controlled by the Linux scheduler.

It is worth noting that the three implementations differ in memory allocation strategy. Namely, for the Goose language stack, data structures managed on per-worker basis are employed. The C++ and Rust tree implementations used two alternative storage strategies. Individually allocated nodes were managed using C++ \texttt{unique\_ptr} and Rust \texttt{Box}. Contiguous arenas were implemented using C++ \texttt{vector} and Rust \texttt{Vec}, with 32-bit child indices. This setup helps to compare the traditional object allocation technique with the compact memory structure allocation strategy. 

In the compact tree representations, each node contained a 32-bit value and two 32-bit child references or indices. The C++ and Rust arenas used child indices, while Goose used nullable self-relative references. The resulting node representation occupied 12 bytes.

In the repeated reclamation benchmark, each Goose worker used one growable pool per round. Each C++ and Rust arena variant used one reserved vector per worker and round. Individually allocated variants created a separate allocation for each node. All trees remained allocated until the LIVE checkpoint.

The workers did not exchange array elements or tree nodes. They communicated results and control information to the main thread. In the tree experiments, Goose used typed message queues. C++ used barriers and shared result records. Rust used barriers and channels. In the dynamic array benchmark, the Goose workers reported partial sums through a queue. C++ workers wrote in separate result slots, while Rust workers returned partial sums through thread handles.

The source code for all tests is available in a public repository \cite{fraczek2026tests}.

\subsection{Benchmark Workloads}

Performance evaluation has covered three workloads to assess allocation effectiveness, data structure traversal efficiency, and memory reclamation. The experiments were performed with different levels of parallelism to measure the scalability.

The dynamic array benchmark processed 64 million 64-bit integers in total. Each worker owned a separate array that contained an equal share of the elements. Each worker filled and traversed its array. The Goose arrays grew dynamically. The C++ and Rust vectors used automatic growth or explicit capacity reservation.

The binary-tree experiments used complete trees of depths 18 and 20.  A tree of depth \(d\) contained \(2^{d+1}-1\) nodes, with the root at depth zero. Each tree was traversed eight times and deterministic checksums were used to verify the computation. Each worker has its own set of data structures; there is no exchange of tree nodes between workers. The repeated-reclamation benchmark performed five consecutive construction, traversal, and release rounds using persistent worker threads. The standalone tree benchmark constructed and released trees individually. The repeated reclamation benchmark used synchronized memory checkpoints before and after release.

We chose complete binary trees because their size and lifetime are predictable. They also allow us to examine the effects of separate node allocations. This is a controlled workload, not a model of every tree application. B-trees and B+ trees have different node layouts and access patterns.

Two scaling scenarios were evaluated. For strong scaling, eight trees were evenly distributed among 1, 2, 4, and 8 workers. For weak scaling, each worker processed one depth-18 tree.

\subsection{Measurement Procedure and Metrics}

An external launcher measured the total process time using \texttt{CLOCK\_MONOTONIC}. Timing started just before process creation and ended after \texttt{wait4()} function returned. It included initialization, computation, synchronization, memory sampling, and termination. The worker timers measured the construction and the traversal together. The release timing started after the LIVE synchronization wait. The wait itself was excluded.

We measured process RSS with \texttt{/proc/self/smaps\_rollup} and VMS with \texttt{/proc/self/statm}. BASE was measured after the workers started and before tree construction. LIVE was measured while the trees remained allocated. ROUND was measured after logical release. FINAL was measured after the workers had terminated. The peak RSS obtained from \texttt{wait4()} function was analyzed separately.

We excluded the final round from post-release RSS comparisons because worker termination could affect the Goose snapshot. Further measurement details are given in Appendix A.

\subsection{Statistical Analysis and Reproducibility}

Initially, the reclamation experiment consisted of 15 repetitions that included five cycles of memory allocation and reclamation for each configuration. This means that each process execution is considered as an independent observation while multiple runs inside a single trial are not recognized. The results obtained during the preliminary smoke-test stage are ignored in further analysis. The repeated-reclamation benchmark used two warm-up executions per configuration. The standalone-tree benchmark used three. These warm-ups were excluded from the analysis. The execution order of the implementations was randomized during the measured runs.

The execution time aggregation includes the median and median absolute deviation (MAD). In order to measure reclamation interval, first the maximal per-round time spent by all workers is identified. For each round, the maximum execution time of the release operation is calculated among all workers. For each process execution, we subtracted its BASE RSS from the LIVE and ROUND measurements. We calculated the median across five LIVE checkpoints and the median across the first four ROUND checkpoints. The reported values were then calculated as medians across 15 independent process executions.

The system performance evaluation is performed on the basis of the ratio of the median execution time. Strong scaling was evaluated with a fixed total workload. The speedup was calculated as \(S(p)=T(1)/T(p)\). Here, \(T(p)\) is the median execution time for \(p\) workers.
Weak scaling was evaluated with a fixed workload per worker. The normalized execution time was calculated as \(N(p)=T(p)/T(1)\). The total workload increased proportionally with the number of workers.
In both cases, execution time refers to the entire process. Differences between performance values are estimated without performing any statistical significance test. The selected comparisons were additionally analyzed using 20000 bootstrap resamples of the independent process-level measurements, with confidence intervals calculated using the percentile method. The resulting 95\% confidence intervals were used to describe the sampling uncertainty.

\section{Experimental Results and Comparison}

\subsection{Speed and strong scaling comparison}

We evaluated parallel execution using dynamic arrays, standalone binary trees, and repeated tree construction and reclamation. The evaluation focused on comparisons between Goose and individually allocated objects and contiguous arenas in C++ and Rust.

\begin{figure}[!ht]
\centering
\begin{minipage}{0.95\textwidth}
    \centering
    \includegraphics[width=\linewidth]{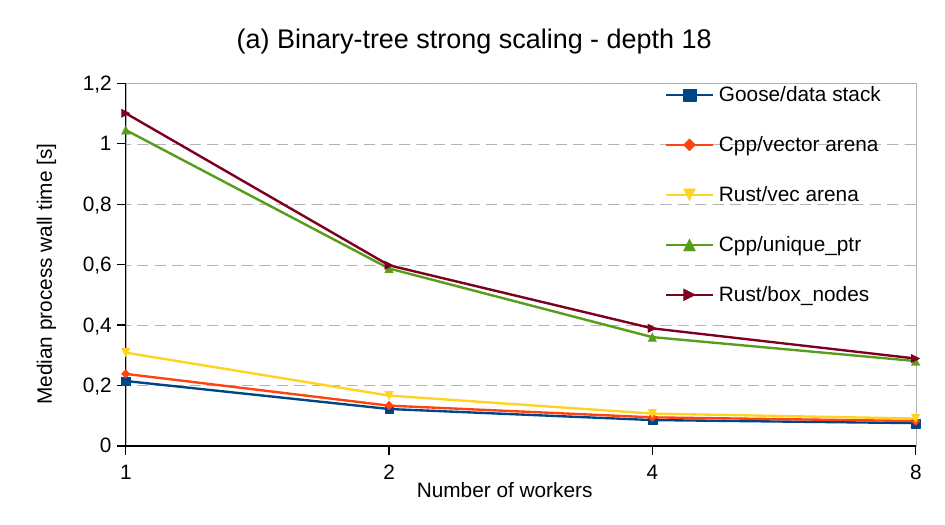}
    \label{rys:fig1a}
\end{minipage}\\[1em]
\begin{minipage}{0.95\textwidth}
    \centering
    \includegraphics[width=\linewidth]{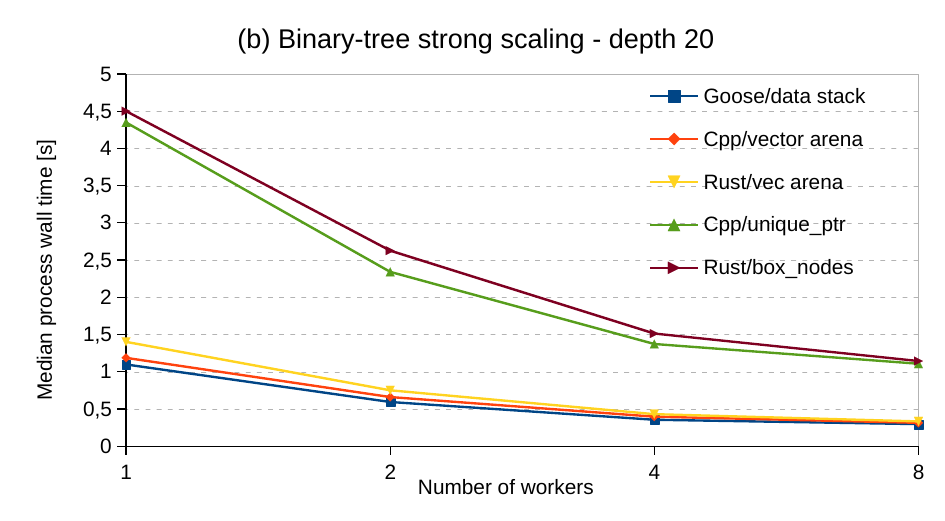}
    \label{rys:fig1b}
\end{minipage}
\caption{Median process wall-clock execution times for binary-tree workloads of (a) depth 18 and (b) depth 20 under strong scaling. Each round processes eight trees, with five rounds and eight traversal passes per round. Results represent 15 independent process executions per configuration.}
\label{rys:fig1}
\end{figure}

The results for the median execution times of the repeated reclamation binary tree benchmarks at tree depths 18 and 20 are shown in Figure \ref{rys:fig1}. Each benchmark run performed eight trees per iteration comprising five construction and reclamation iterations, as well as eight traversals per iteration. The data were collected by performing fifteen runs of each process implementation. Goose produced the fastest median execution times among all five implementations tested.

The execution time at a depth of 20 by a single worker was 1098 ms for Goose, compared to 1186 ms and 1401 ms for the C++ vector arena and the Rust vector arena, respectively. With eight workers, execution times went down to 296 ms, 313 ms, and 332 ms, respectively. Using eight workers, Goose managed to decrease the median execution time by 5.62\% compared to the optimized C++ arena and by 10.95\% compared to the Rust arena. The 95\% bootstrap percentile confidence intervals were 4.90-6.33\% and 10.31-11.20\%, respectively. These estimates were based on 15 independent process executions per configuration.

Performance was much lower when objects were allocated individually. For depth 20 and eight workers, the C++ \texttt{unique\_ptr} implementation had a runtime of 1106 ms and the Rust \texttt{Box} had a runtime of 1144 ms, which means that these took about 3.74 times and 3.87 times the runtime of Goose, respectively. The relatively minor differences between Goose and the optimized arenas indicate that the allocation granularity and representation matter greatly for the performance difference.

Strong-scaling speedups were calculated from the execution times, which are shown in Figure \ref{rys:fig1}. Each speedup was measured relative to the corresponding execution time of a single-worker. At the depth of 20, Goose achieved a speedup of 3.71 with eight workers, as compared to 3.78 for the C++ arena and 4.22 for the Rust arena. Goose achieved the shortest execution time. However, its speedup was lower than that of the C++ and Rust arenas. Therefore, Goose did not achieve the highest strong-scaling efficiency.

The standalone-tree benchmark used a different execution strategy. Each worker constructed, traversed, and released its trees one at a time. In contrast, the repeated-reclamation benchmark kept all assigned trees alive until the LIVE checkpoint. Therefore, the execution times of these two benchmarks are not directly comparable. In the case of depth-20 binary trees with eight workers, Goose took 62 ms, while the C++ arena and the Rust arena took 67 ms and 69 ms, respectively. The results followed the ordering obtained from the repeated reclamation benchmarks. The median execution times for the dynamic-array benchmark are summarized in Table \ref{tab:implementation_results}.

In the case of the dynamic-array benchmark, a better-balanced comparison was obtained. Processing 64 million 64-bit integers with four workers took 40.05 ms with Goose, while it took 39.53 ms with the C++ vector and 40.76 ms with the Rust vector, respectively. For eight workers, it took 43 ms with Goose, while 42 ms with the C++ vector and 46 ms with the Rust vector. Hence, Goose was still competitive, but did not always outperform optimized vectors, highlighting the importance of capacity and workload considerations.

The analysis of strong scaling for repeated reclamation covered 40 configurations. Their relative median absolute deviations ranged from approximately 0.21\% to 3.97\%. Generally, this suggests low variability between runs. The overall execution time results confirm the competitiveness of Goose in the considered cases, while different storage representation, allocation strategies, compiler optimizations, and synchronization make it impossible to attribute any advantages to language features alone.

\begin{table}[!ht]
\centering
\caption{Dynamic-array benchmark: median whole-process execution time (ms) for 64 million 64-bit integers.}
\label{tab:implementation_results}
  \begin{tabular}{|l|r|r|r|r|}
    \hline
    Implementation & 1 worker & 2 workers & 4 workers & 8 workers \\\hline
    Goose stack growth  & 92.86  & 53.35 & 40.05 & 42.86 \\\hline
    C++ vector growth   & 143.28 & 99.45 & 80.50 & 81.78 \\\hline
    C++ vector reserve  & 92.69  & 52.97 & 39.53 & 42.09 \\\hline
    Rust vector growth  & 92.41  & 57.42 & 44.31 & 48.26 \\\hline
    Rust vector reserve & 88.00  & 53.18 & 40.76 & 46.47 \\\hline
  \end{tabular}
\end{table}

\subsection{Memory Consumption}

\begin{figure}[!ht]
\centering
\begin{minipage}{0.48\textwidth}
    \centering
    \includegraphics[width=\linewidth]{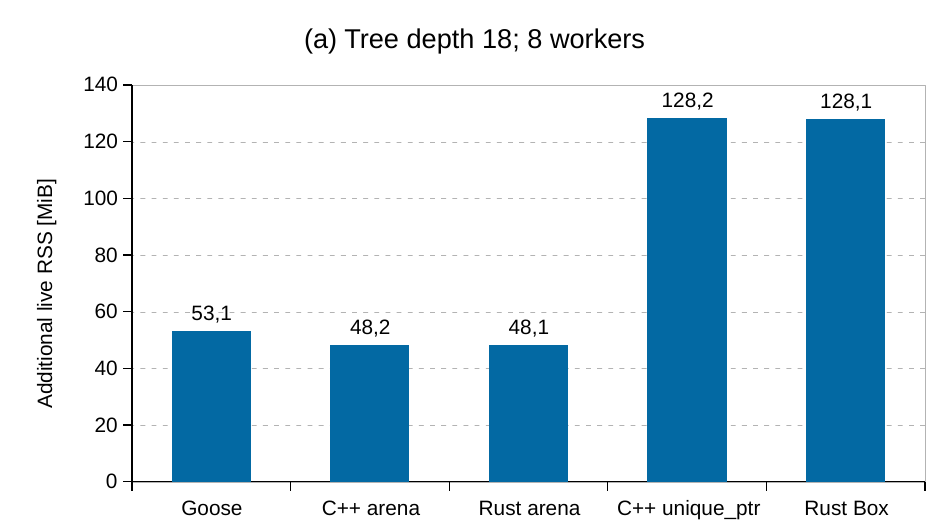}
    \label{rys:fig2a}
\end{minipage}\hfill
\begin{minipage}{0.48\textwidth}
    \centering
    \includegraphics[width=\linewidth]{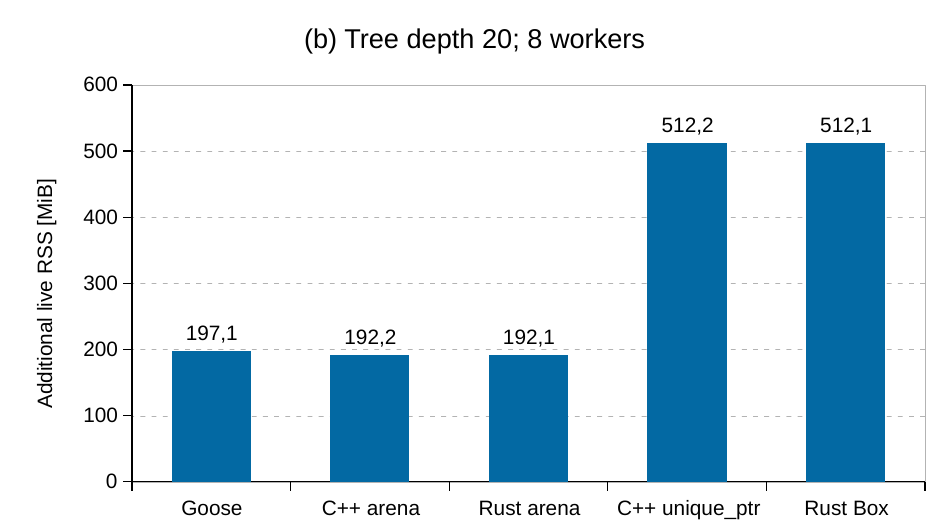}
    \label{rys:fig2b}
\end{minipage}
\caption{Additional resident set size (RSS) measured at synchronized LIVE checkpoints for binary tree workloads with eight workers: (a) depth 18 and (b) depth 20.}
\label{rys:fig2}
\end{figure}

Memory usage of the program was assessed based on synchronized process-level metrics collected during the multiple memory reclaim procedure. The results shown in Figure \ref{rys:fig2} represent the additional size of the resident set (RSS) at the LIVE checkpoints, calculated by subtracting the corresponding BASE RSS measurements.

For depth-20 trees with eight workers, the Goose implementation exhibited 197 MiB of additional process RSS relative to BASE. The corresponding values were 192 MiB for the C++ arena and 192 MiB for the Rust arena, as shown in Figure \ref{rys:fig2}. In contrast, individually allocated C++ \texttt{unique\_ptr} and Rust \texttt{Box} nodes take about 512 MiB and 512 MiB, which is approximately 2.67 times the additional RSS of the corresponding arena-based implementations. For depth-18 trees with eight workers, Goose used approximately 53 MiB of additional RSS. The C++ and Rust arenas used approximately 48 MiB. The individually allocated implementations used approximately 128 MiB.

\begin{figure}[!ht]
\centering
\begin{minipage}{0.9\textwidth}
    \centering
    \includegraphics[width=\linewidth]{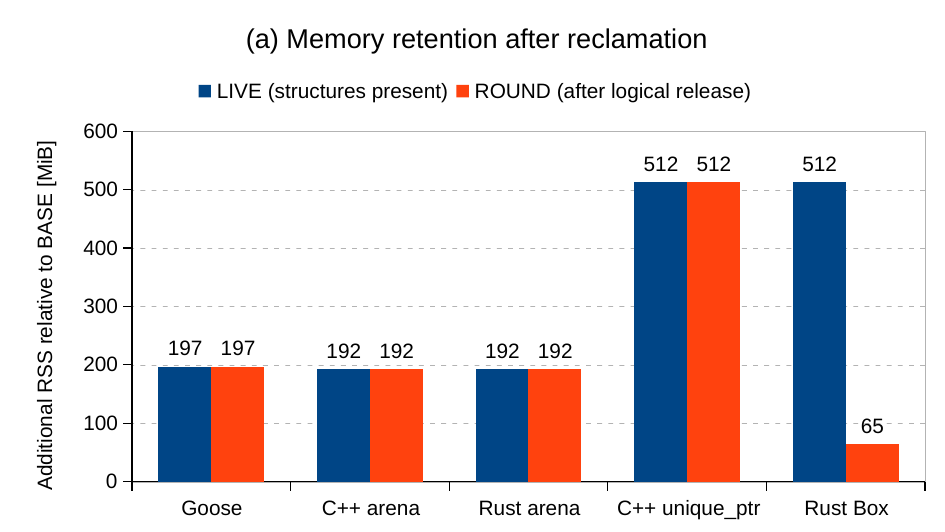}
    \label{rys:fig3a}
\end{minipage}\\[1em]
\begin{minipage}{0.9\textwidth}
    \centering
    \includegraphics[width=\linewidth]{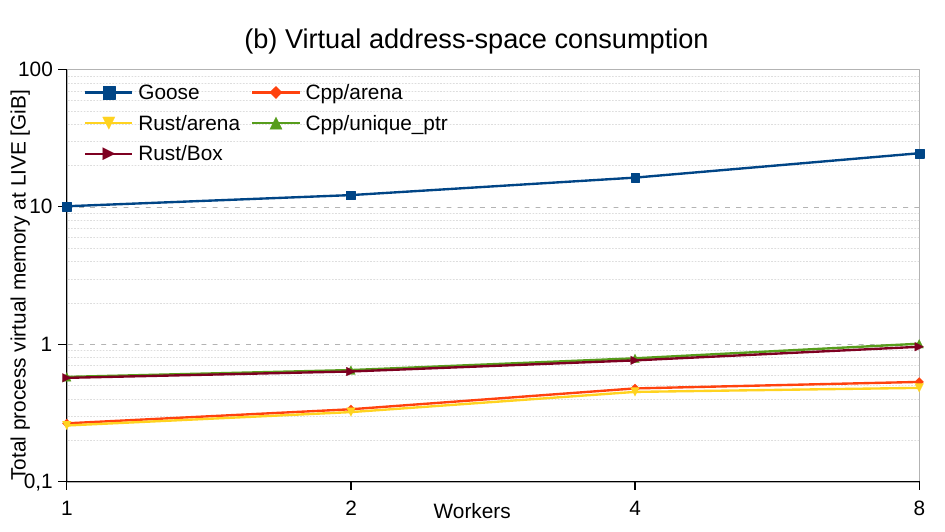}
    \label{rys:fig3b}
\end{minipage}
\caption{Memory-management behavior of the evaluated implementations: (a) baseline-adjusted resident set size before and after logical reclamation for depth-20 binary trees with eight workers; (b) total virtual memory size at the LIVE checkpoint as a function of worker count. Results represent medians from 15 independent executions. Post-release RSS values exclude the final round.}
\label{rys:fig3}
\end{figure}

Figure \ref{rys:fig3}a compares process RSS before and after logical reclamation. The analysis includes the first four rounds. For depth-20 trees with eight workers, Goose retained approximately 197 MiB of additional RSS above baseline after release. The C++ and Rust arenas each retained approximately 192 MiB, while C++ \texttt{unique\_ptr} retained approximately 512 MiB. In contrast, Rust \texttt{Box} had approximately 65 MiB of additional RSS after release. With one worker, the arena implementations reduced RSS to values close to the baseline.

The virtual address space consumption measurements helped identify another interesting distinction between the programming languages under investigation. Figure \ref{rys:fig3}b shows the total VMS of the process for different counts of workers. For Goose, the VMS ranged from approximately 10.08 GiB with one worker to 24.52 GiB with eight workers. These large VMS values were already present at the BASE checkpoint, before tree construction. Other languages demonstrated much lower values ranging from approximately 0.48 to 1.01 GiB in the case of eight worker configuration.

Goose exhibited a resident memory consumption close to that of the contiguous arenas. However, it reserved substantially more virtual address space. These reservations did not correspond to an equivalent amount of physically resident memory. From the results presented, it is possible to draw conclusions about the three different aspects - object representation, resident memory consumption, and virtual address space requirements.

\subsection{Reclamation Performance}

The evaluation of the performance of the reclamation algorithm involved measuring the time required to perform storage release operations. We note that the corrected instrumentation does not take into account the waiting times incurred in the LIVE synchronization barrier phase. In each round, the maximum time that workers required to complete release operations was recorded. The median statistics from five rounds were collected separately for each independent process run repeated 15 times.

\begin{figure}[!ht]
\centering
\includegraphics[scale=.8]{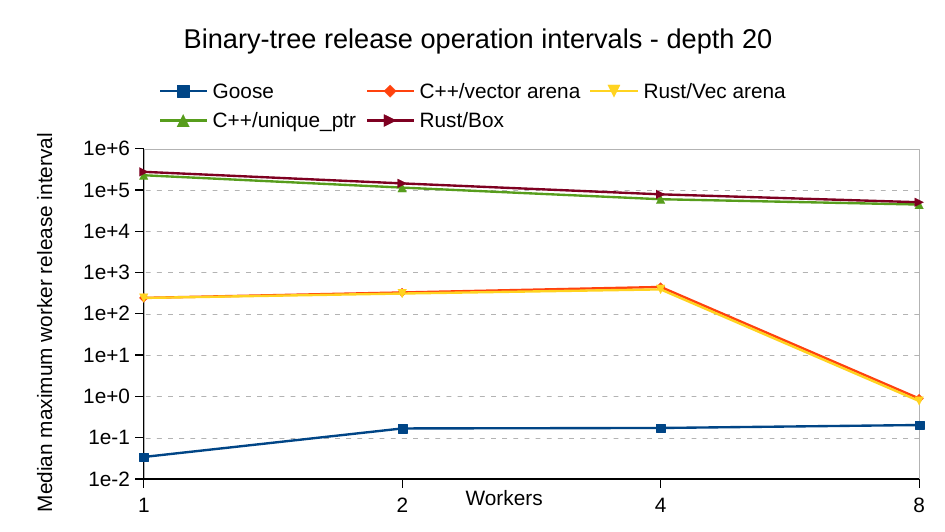}
\caption{Median maximum worker release intervals in microseconds for depth-20 trees. The vertical axis is logarithmic.}
\label{rys:fig4}
\end{figure}

The release intervals were calculated for each depth of the tree separately. Figure \ref{rys:fig4} shows the release intervals for depth-20 trees. The median maximum worker release interval for C++ \texttt{unique\_ptr} and the Rust \texttt{Box} is 44 and 51 ms, respectively, with 8 workers involved in the calculations. Meanwhile, the release interval for the contiguous C++ arena and Rust arena was found to be around 0.9 and 0.8 $\mu$s, respectively. The Goose library scope exit latency was measured to be equal to approximately 0.2 $\mu$s. The qualitative comparison results for depth-18 trees were analogous.

The release interval also depended on the number of workers involved. When the depth was 20 and only one worker was active, the release interval of the C++ and Rust arena allocators was approximately 242 $\mu$s. When there were eight workers working simultaneously, the release interval decreased dramatically and became around 1 microsecond. The total workload remained constant across worker configurations. However, the workload assigned to each worker decreased as the worker count increased. Consequently, the amount of memory released by each worker also changed. These observations corresponded to differences in post-release RSS.

The release operations differ between implementations. Goose restores a compiler-maintained stack position. The C++ and Rust arena variants release their vector storage. Individually allocated variants recursively destroy tree nodes and release their allocations. Goose release intervals were extremely short. Their measured values may be affected by the overhead and resolution of clock calls. This overhead was not calibrated separately. Therefore, the results should not be interpreted as exact measurements of physical memory reclamation.

These experiments show that collective release can be inexpensive when objects have compatible lifetimes. However, the measured release operations differ between implementations. The shortest Goose intervals are also sensitive to clock-measurement overhead. The results of the study demonstrate that the evaluation of the release operation must be supplemented with memory retention estimates.

\subsection{Weak-Scaling Behavior}

\begin{figure}[!ht]
\centering
\includegraphics[scale=.75]{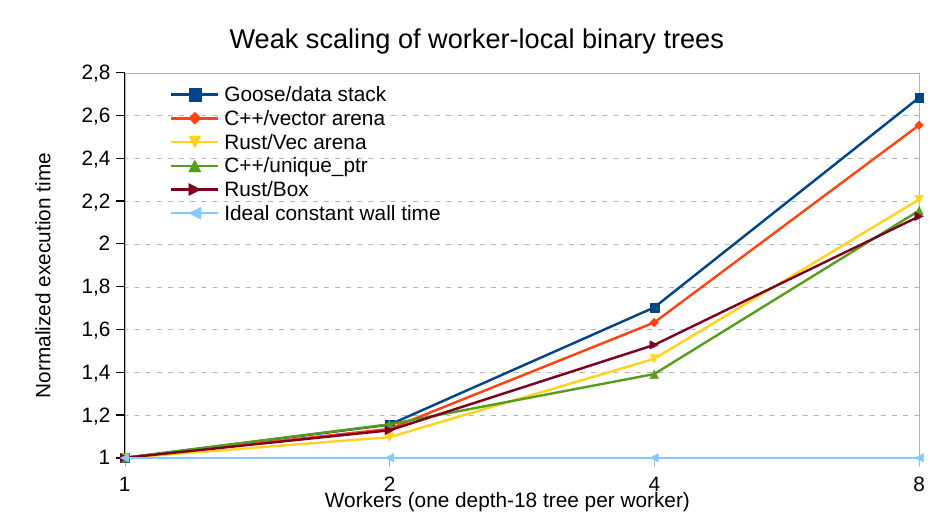}
\caption{Normalized median execution time under weak scaling for depth-18 binary trees, with one tree per worker. Each configuration comprised five rounds and eight traversal passes per round. Results represent medians from 15 independent process executions. Ideal weak scaling corresponds to a constant normalized execution time of 1.}
\label{rys:fig5}
\end{figure}

The tests have been conducted using complete binary trees of depth equal to 18, with each worker being responsible for creating one tree. Therefore, the overall amount of work increased linearly depending on the number of workers. In each experiment, five cycles consisting of eight traversals of constructed and reclaimed binary trees have been conducted. The tests have been carried out with 1, 2, 4, and 8 workers, each setup being repeated independently fifteen times. We assessed weak scaling using normalized execution time $T(p)/T(1)$. A ratio of 1 represents ideal weak scaling. Smaller ratios are better. We also calculated the weak scaling efficiency as $T(1)/T(p)$. Figure \ref{rys:fig5} shows the normalized execution times for five implementations tested above. The implementation based on Goose demonstrated the shortest absolute execution time for all numbers of workers. With 1 worker, the execution time is equal to 28 ms, while in the case of 2, 4, and 8 workers, it is equal to 32 ms, 48 ms, and 75 ms, respectively. The arena allocator implemented in C++ increases the execution time from 32 to 82 ms, whereas the arena allocator implemented in Rust goes up to 90 ms for 8 workers. With eight workers, the individually allocated C++ and Rust tree implementations required 283 ms and 287 ms, respectively. In terms of relative performance degradation, Goose showed the highest relative growth, which was equal to 2.68. The relative growth in execution time in other cases equals 2.55, 2.21, 2.16 and 2.13 for the C++ arena, the Rust arena, the C++ \texttt{unique\_ptr} and the Rust \texttt{Box}, respectively. Weak-scaling efficiencies in percent have equaled 37.3\%, 39.2\%, 45.3\%, 46.4\%, and 47.0\%. This means that even those implementations with smaller absolute execution times did not demonstrate better scaling efficiencies relative to other implementations.

The uncertainty of the sampling was evaluated using 20000 bootstrap resamples of independent process-level measurements, with confidence intervals estimated using the percentile method. The resulting 95\% confidence intervals describe the precision of the normalized execution-time estimates under the tested conditions. Figure \ref{rys:fig5} shows the normalized execution times of one to eight workers. The 95\% confidence interval for normalized execution time for Goose ranges from 2.65 to 2.78, whereas in the case of the C++ arena, the interval equals 2.46 to 2.63 and for the Rust arena, the interval ranges from 2.14 to 2.30. Presented figures depict confidence intervals caused by the sampling variability.

A marked increase in execution time was observed when moving from four to eight workers in the weak-scaling experiments. Taking into account the fact that there are 4 cores available and 8 logical processors provided by a CPU, the last case implies simultaneous multithreading mode. At the moment, no information about the separation of influence caused by simultaneous multithreading, memory bandwidth limitation, and synchronization overhead is presented in the current analysis.

Table \ref{tab:implementation_scaling} summarizes weak-scaling execution times and normalized eight-to-one-worker ratios.

\begin{table}[!ht]
\centering
\caption{Weak-scaling performance across implementations. Execution times are reported in milliseconds (ms). The T(8)/T(1) ratio is dimensionless.}
\label{tab:implementation_scaling}
  \begin{tabular}{|l|r|r|r|r|r|}
    \hline
    Implementation & 1 worker & 2 workers & 4 workers & 8 workers & T(8)/T(1) \\\hline
    Goose            & 27.99  & 32.31  & 47.64  & 75.08  & 2.68 \\\hline
    C++ vector arena & 32.13  & 36.48  & 52.45  & 82.07  & 2.55 \\\hline
    Rust Vec arena   & 40.71  & 44.63  & 59.53  & 89.82  & 2.21 \\\hline
    C++ unique\_ptr  & 131.19 & 151.72 & 182.52 & 282.72 & 2.16 \\\hline
    Rust Box         & 135.02 & 152.37 & 206.20 & 287.33 & 2.13 \\\hline
  \end{tabular}
\end{table}
All five implementations showed increasing execution times as the worker count increased. None achieved ideal weak scaling.

\section{Discussion}

\subsection{Discussion of Main Findings}

Based on experimental results obtained on binary trees, managed storage performance has been demonstrated to be highly dependent on workload parameters, storage representations, and parallel computation features. In terms of execution time, Goose achieves the minimum values compared to C++ and Rust arena implementations. However, it shows a substantial performance advantage over individually allocated nodes. Previous research has also shown that allocation policies and the layout of the data-structure influence performance \cite{DBLP:conf/oopsla/BergerZM02} \cite{DBLP:conf/pldi/LattnerA05}. This supports the use of optimized arena implementations as comparison baselines.

Regarding the depth-20 test case with eight workers, Goose decreases the median execution time of the benchmark by 5.62\% relative to the C++ arena and by 10.95\% relative to the Rust arena. Having considered the 95\% bootstrap percentile confidence intervals obtained, one may conclude that there is a positive difference in favor of Goose, especially taking into account the variability due to the small number of process executions. However, these findings should be interpreted with caution, as these benefits depend on particular workloads. Experiments carried out on the single hardware architecture cannot reveal if these benefits generalize to other architectures.

The experimental results show that storage managed by the compiler on a per-worker basis demonstrates competitiveness in relation to data structures. Specifically, in experiments with binary trees, Goose was faster than the highly optimized C++ and Rust arenas. Nevertheless, such a superiority was not observed in the dynamic array benchmarks.

Concerning the memory consumption metrics, Goose shows comparable values of resident set sizes with contiguous arenas and considerably smaller than those in the case of individually allocated nodes. Significantly, the observed difference is mainly due to the data storage strategies considered and cannot be attributed exclusively to allocation methods. In particular, compact arena-based nodes avoid overhead costs related to separate memory allocations. Goose exhibited a substantially larger process VMS than the other implementations. With eight workers, the VMS was approximately 24.52 GiB. This value was already present before tree construction. The reservation of virtual addresses should be considered in terms of resource usage.

Reclamation experiments showed substantial differences in the duration of the release-operation. Goose had the shortest measured intervals. The C++ and Rust arena implementations also supported bulk release, but used different mechanisms. These timings should be interpreted carefully because the shortest intervals were sensitive to clock-measurement overhead. Moreover, released memory often remained resident.

Parallel scaling introduced another factor influencing the overall performance of the benchmarks in the experiments. Goose achieved the smallest absolute execution times. However, it was not characterized by consistent maximal efficiency during a strong parallel scaling of the workload. Regarding weak scaling, Goose showed the highest normalized execution time (2.68) of the five ratios, where a lower is better. The ratio for the C++ arena benchmark was 2.55 and the normalized execution time of the Rust arena was 2.21. The eight-worker configuration allowed for the use of simultaneous multithreading on the four-core processor.

This study extends previous Goose benchmarks \cite{bakker2026goosebench} with measurements of parallel performance and memory reclamation. The results indicate that compiler-managed worker-local storage can be competitive for the tested workloads. However, its evaluation should also consider resident memory, virtual address space, and scaling efficiency.

\subsection{Limitations}

The implementations performed equivalent computations, verified by node counts and checksums. However, their data layouts, allocation methods, and synchronization mechanisms differed. The C++ and Rust arenas used compact nodes with 32-bit child indices. The other C++ and Rust tree variants used pointers and separate node allocations. Thus, timing differences cannot be attributed to language design or allocation overhead alone.

We used complete binary trees to examine the cost of allocating individual nodes. These trees do not represent every application. B-trees and B+ trees could produce different results. We did not measure cache misses, cache hit rates, or memory access latency. Differences in locality may therefore contribute to the observed execution times.

The entire process execution time consists of initialization, computation, synchronization, and termination. RSS and VMS were sampled at checkpoints rather than continuously monitored. The release intervals measured implementation-specific operations, not the return of physical memory to the operating system. Very short Goose scope-exit intervals may be affected by clock measurement overhead and resolution, which were not calibrated.

The total process execution times also included the overhead of collecting memory checkpoints. This overhead was not measured separately. It may have affected the results, particularly for short workloads.

In the final Goose round, worker threads could terminate during the ROUND memory snapshot. Therefore, only the first four rounds were included in post-release RSS comparisons.

All experiments were conducted on a workstation with an Intel Core i3-14100 processor. Process affinity was restricted to selected logical CPUs, but individual workers were not tied to specific cores. Scheduling, simultaneous multithreading, CPU frequency variations, and resource contention may have affected the results.

The main experiments used 15 independent process executions per configuration. The standalone depth-18 pilot experiments used five executions. These measurements support descriptive statistics and bootstrap confidence intervals, but do not establish performance across other systems. The benchmarks used synthetic worker-local arrays and binary trees. Shared data structures, irregular workloads, and other hardware architectures were not evaluated.

\section{Conclusions}

This study examined the performance and memory behavior of Goose v2026.3 in parallel workloads. We compared Goose with C++ and Rust implementations using individual object allocations and optimized contiguous arenas. The experiments covered dynamic arrays and binary trees. We evaluated execution time, memory consumption, release-operation duration, and scaling.

In the repeated reclamation benchmark at depth 20 with eight workers, Goose achieved a median execution time 5.62\% lower than the C++ arena and 10.95\% lower than the Rust arena. Its additional RSS at the LIVE checkpoint was comparable to that of both arenas and substantially lower than that of individually allocated trees. However, Goose exhibited much larger process VMS, reaching approximately 24.52 GiB. Goose also showed very short scope-exit intervals, although clock measurement overhead limits their precision. Logical reclamation did not always reduce process RSS.

Despite shorter execution times in the repeated reclamation experiments, Goose showed lower weak scaling efficiency than the C++ and Rust arenas. The results of the dynamic array tests were also mixed. These findings indicate that the benefits of worker-local storage managed by the compiler depend on the workload and the performance metric. They also highlight the importance of distinguishing logical memory reclamation from changes in resident memory.

The experiments were limited to synthetic worker workloads on a single workstation. Differences in data representation and synchronization also affect the comparisons. Future work should examine larger multicore systems and applications with shared or irregular data structures. It should also investigate B-trees, B+ trees, and cache behavior. Another useful experiment would be to compare C++ and Rust arenas that retain their capacity between rounds. This could help assess the effect of reusing allocated storage. In general, evaluating parallel memory management requires considering execution time, resident memory, virtual address space, and reclamation behavior together.

\bibliographystyle{cs-agh}
\bibliography{bibliography}

@inproceedings{DBLP:conf/asplos/BergerMBW00,
  author       = {Emery D. Berger and
                  Kathryn S. McKinley and
                  Robert D. Blumofe and
                  Paul R. Wilson},
  editor       = {Larry Rudolph and
                  Anoop Gupta},
  title        = {Hoard: {A} Scalable Memory Allocator for Multithreaded Applications},
  booktitle    = {{ASPLOS-IX} Proceedings of the 9th International Conference on Architectural
                  Support for Programming Languages and Operating Systems, Cambridge,
                  MA, USA, November 12-15, 2000},
  pages        = {117--128},
  publisher    = {{ACM} Press},
  year         = {2000},
  url          = {https://doi.org/10.1145/378993.379232},
  doi          = {10.1145/378993.379232},
  bibsource    = {dblp computer science bibliography, https://dblp.org}
}

@inproceedings{DBLP:conf/oopsla/BergerZM02,
  author       = {Emery D. Berger and
                  Benjamin G. Zorn and
                  Kathryn S. McKinley},
  editor       = {Mamdouh Ibrahim and
                  Satoshi Matsuoka},
  title        = {Reconsidering custom memory allocation},
  booktitle    = {Proceedings of the 2002 {ACM} {SIGPLAN} Conference on Object-Oriented
                  Programming Systems, Languages and Applications, {OOPSLA} 2002, Seattle,
                  Washington, USA, November 4-8, 2002},
  pages        = {1--12},
  publisher    = {{ACM}},
  year         = {2002},
  url          = {https://doi.org/10.1145/582419.582421},
  doi          = {10.1145/582419.582421},
  bibsource    = {dblp computer science bibliography, https://dblp.org}
}

@book{DBLP:books/daglib/0019344,
  author       = {Bjarne Stroustrup},
  title        = {The {C++} programming language - special edition {(3.} ed.)},
  publisher    = {Addison-Wesley},
  year         = {2000},
  isbn         = {978-0-201-70073-2},
  bibsource    = {dblp computer science bibliography, https://dblp.org}
}

@article{DBLP:journals/pacmpl/0002JKD18,
  author       = {Ralf Jung and
                  Jacques{-}Henri Jourdan and
                  Robbert Krebbers and
                  Derek Dreyer},
  title        = {RustBelt: securing the foundations of the rust programming language},
  journal      = {Proc. {ACM} Program. Lang.},
  volume       = {2},
  number       = {{POPL}},
  pages        = {66:1--66:34},
  year         = {2018},
  url          = {https://doi.org/10.1145/3158154},
  doi          = {10.1145/3158154},
  bibsource    = {dblp computer science bibliography, https://dblp.org}
}

@misc{Wouters2026,
  author = {van Oortmerssen, W.},
  title = {The Goose Programming Language},
  year = {2026},
  publisher = {GitHub},
  journal = {GitHub repository},
  howpublished = {\url{https://github.com/aardappel/goose}},
  commit = {8511ae34bdc9aabf77baaab80b4d659ea87835e2}
}

@article{DBLP:journals/iandc/TofteT97,
  author       = {Mads Tofte and
                  Jean{-}Pierre Talpin},
  title        = {Region-based Memory Management},
  journal      = {Inf. Comput.},
  volume       = {132},
  number       = {2},
  pages        = {109--176},
  year         = {1997},
  url          = {https://doi.org/10.1006/inco.1996.2613},
  doi          = {10.1006/INCO.1996.2613},
  bibsource    = {dblp computer science bibliography, https://dblp.org}
}

@inproceedings{DBLP:conf/pldi/GrossmanMJHWC02,
  author       = {Dan Grossman and
                  J. Gregory Morrisett and
                  Trevor Jim and
                  Michael W. Hicks and
                  Yanling Wang and
                  James Cheney},
  editor       = {Jens Knoop and
                  Laurie J. Hendren},
  title        = {Region-Based Memory Management in Cyclone},
  booktitle    = {Proceedings of the 2002 {ACM} {SIGPLAN} Conference on Programming
                  Language Design and Implementation (PLDI), Berlin, Germany, June 17-19,
                  2002},
  pages        = {282--293},
  publisher    = {{ACM}},
  year         = {2002},
  url          = {https://doi.org/10.1145/512529.512563},
  doi          = {10.1145/512529.512563},
  bibsource    = {dblp computer science bibliography, https://dblp.org}
}

@inproceedings{DBLP:conf/pldi/LattnerA05,
  author       = {Chris Lattner and
                  Vikram S. Adve},
  editor       = {Vivek Sarkar and
                  Mary W. Hall},
  title        = {Automatic pool allocation: improving performance by controlling data
                  structure layout in the heap},
  booktitle    = {Proceedings of the {ACM} {SIGPLAN} 2005 Conference on Programming
                  Language Design and Implementation, Chicago, IL, USA, June 12-15,
                  2005},
  pages        = {129--142},
  publisher    = {{ACM}},
  year         = {2005},
  url          = {https://doi.org/10.1145/1065010.1065027},
  doi          = {10.1145/1065010.1065027},
  bibsource    = {dblp computer science bibliography, https://dblp.org}
}

@inproceedings{DBLP:conf/aplas/LeijenZM19,
  author       = {Daan Leijen and
                  Benjamin Zorn and
                  Leonardo de Moura},
  editor       = {Anthony Widjaja Lin},
  title        = {Mimalloc: Free List Sharding in Action},
  booktitle    = {Programming Languages and Systems - 17th Asian Symposium, {APLAS}
                  2019, Nusa Dua, Bali, Indonesia, December 1-4, 2019, Proceedings},
  series       = {Lecture Notes in Computer Science},
  volume       = {11893},
  pages        = {244--265},
  publisher    = {Springer},
  year         = {2019},
  url          = {https://doi.org/10.1007/978-3-030-34175-6\_13},
  doi          = {10.1007/978-3-030-34175-6\_13},
  bibsource    = {dblp computer science bibliography, https://dblp.org}
}

@article{DBLP:journals/tpds/Michael04,
  author       = {Maged M. Michael},
  title        = {Hazard Pointers: Safe Memory Reclamation for Lock-Free Objects},
  journal      = {{IEEE} Trans. Parallel Distributed Syst.},
  volume       = {15},
  number       = {6},
  pages        = {491--504},
  year         = {2004},
  url          = {https://doi.org/10.1109/TPDS.2004.8},
  doi          = {10.1109/TPDS.2004.8},
  bibsource    = {dblp computer science bibliography, https://dblp.org}
}

@inproceedings{DBLP:conf/asplos/ZhouGVK0KMDR24,
  author       = {Zhuangzhuang Zhou and
                  Vaibhav Gogte and
                  Nilay Vaish and
                  Chris Kennelly and
                  Patrick Xia and
                  Svilen Kanev and
                  Tipp Moseley and
                  Christina Delimitrou and
                  Parthasarathy Ranganathan},
  editor       = {Rajiv Gupta and
                  Nael B. Abu{-}Ghazaleh and
                  Madan Musuvathi and
                  Dan Tsafrir},
  title        = {Characterizing a Memory Allocator at Warehouse Scale},
  booktitle    = {Proceedings of the 29th {ACM} International Conference on Architectural
                  Support for Programming Languages and Operating Systems, Volume 3,
                  {ASPLOS} 2024, La Jolla, CA, USA, 27 April 2024- 1 May 2024},
  pages        = {192--206},
  publisher    = {{ACM}},
  year         = {2024},
  url          = {https://doi.org/10.1145/3620666.3651350},
  doi          = {10.1145/3620666.3651350},
  bibsource    = {dblp computer science bibliography, https://dblp.org}
}

@article{DBLP:journals/pacmpl/HughesT25,
  author       = {Jacob Hughes and
                  Laurence Tratt},
  title        = {Garbage Collection for Rust: The Finalizer Frontier},
  journal      = {Proc. {ACM} Program. Lang.},
  volume       = {9},
  number       = {{OOPSLA2}},
  pages        = {3588--3614},
  year         = {2025},
  url          = {https://doi.org/10.1145/3763179},
  doi          = {10.1145/3763179},
  bibsource    = {dblp computer science bibliography, https://dblp.org}
}

@article{DBLP:journals/pacmpl/HeJBR26,
  author       = {Siyuan He and
                  Songlin Jia and
                  Yuyan Bao and
                  Tiark Rompf},
  title        = {When Lifetimes Liberate: {A} Type System for Arenas with Higher-Order
                  Reachability Tracking},
  journal      = {Proc. {ACM} Program. Lang.},
  volume       = {10},
  number       = {{OOPSLA1}},
  pages        = {1486--1513},
  year         = {2026},
  url          = {https://doi.org/10.1145/3798254},
  doi          = {10.1145/3798254},
  bibsource    = {dblp computer science bibliography, https://dblp.org}
}

@misc{bakker2026goosebench,
  author       = {Wouter van Oortmerssen},
  title        = {{Goose Benchmark Results}},
  url          = {https://aardappel.github.io/goose/bench/results.html},
  year         = {2026},
  month        = {October},
  note         = {Online; accessed 2 November 2026}
}

@misc{fraczek2026tests,
  author       = {Rafał Frączek},
  title        = {{Test sources}},
  url          = {https://bitbucket.org/rafalfr/goosetests/commits/e3454aa9df57492371c89561a46e596821c3c805},
  year         = {2026},
  month        = {November},
  note         = {Commit e3454aa; online; accessed 2 November 2026}
}

@inproceedings{Evans2006jemalloc,
  author    = {Jason Evans},
  title     = {A Scalable Concurrent malloc(3) Implementation for FreeBSD},
  booktitle = {BSDCan 2006},
  year      = {2006},
  url       = {https://papers.freebsd.org/2006/bsdcan/evans-jemalloc/}
}

\appendix
\section{Measurement Details}

The external launcher called the function \texttt{clock\_gettime(CLOCK\_MONOTONIC)} immediately before \texttt{fork()}. It started the benchmark in a child process and used \texttt{wait4()} function to detect completion. It reads \texttt{ru\_maxrss} from the resource statistics of the child process. This records the maximum process RSS, not the number of allocations.

In the repeated reclamation benchmark, workers finished the construction and traversal before the main thread took each LIVE snapshot. The trees remained allocated during the snapshot. After synchronization, each worker timed its own storage release operation. The next ROUND snapshot was taken only after all workers had completed the release. Worker timers excluded synchronization waiting, but total process time included it.

The shared helper reads the Rss: field from \texttt{/proc/self/smaps\_rollup}. It also reads the virtual memory size from \texttt{/proc/self/statm}. Both measurements describe the whole process. Resident memory can remain allocated to the process after logical release and may be reused. No explicit memory trimming was performed. We kept the maximum RSS reported by \texttt{wait4()} function separate from the sampled RSS measurements. The fifth ROUND sample was excluded from post-release comparisons because Goose workers could terminate around this checkpoint.

\vspace{1.5em}
\begin{center}
\small\textit{The paper has been accepted for publication in Computer Science journal: \url{http://journals.agh.edu.pl/csci}}
\end{center}

\end{document}